\documentclass{article}
\pdfoutput=1

\usepackage{arxiv}

\usepackage[utf8]{inputenc} 
\usepackage[T1]{fontenc}    
\usepackage{hyperref}       
\usepackage{url}            
\usepackage{booktabs}       
\usepackage{amsfonts}       
\usepackage{nicefrac}       
\usepackage{microtype}      
\usepackage{lipsum}		
\usepackage{graphicx}
\usepackage{natbib}
\usepackage{doi}

\title{GTFS2STN: Analyzing GTFS Transit Data By Generating Spatiotemporal Transit Network}

\author{ \href{https://orcid.org/0000-0002-6540-4108}{\includegraphics[scale=0.06]{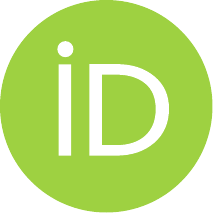}\hspace{1mm}Diyi Liu}\thanks{the GTFS2STN application (beta version) is available at: \url{https://gtfs2stn.streamlit.app}; the corresponding source code is available at: \url{https://github.com/thefriedbee/GTFS2STN}; } \\
	Department of Civil and Environmental Engineering\\
	University of Tennessee, Knoxville\\
	Knoxville, Tennessee, USA \\
	\texttt{dliu27@vols.utk.edu} \\
	\And
	\href{https://orcid.org/0000-0002-3489-3746}{\includegraphics[scale=0.06]{orcid.pdf}\hspace{1mm}Jing Guo} \\
	Department of Civil and Environmental Engineering\\
	Changsha University of Science \& Technology\\
	Changsha, China \\
	\texttt{jingguo97@outlook.com} \\
	\And
    \href{https://orcid.org/0000-0003-0971-9540}{\includegraphics[scale=0.06]{orcid.pdf}\hspace{1mm}Yangsong Gu} \\
	Department of Civil and Environmental Engineering\\
	University of Tennessee, Knoxville\\
	Knoxville, Tennessee, USA \\
	\texttt{ygu17@vols.utk.edu}
    \And
    \href{https://orcid.org/0000-0002-1381-8739}{\includegraphics[scale=0.06]{orcid.pdf}\hspace{1mm}Meredith King}\\
	Department of Civil and Environmental Engineering\\
	University of Tennessee, Knoxville\\
	Knoxville, Tennessee, USA \\
	\texttt{meredithaking@outlook.com}
    \And
    \href{https://orcid.org/0000-0002-1381-1254}{\includegraphics[scale=0.06]{orcid.pdf}\hspace{1mm}Lee D. Han} \\
	Department of Civil and Environmental Engineering\\
	University of Tennessee, Knoxville\\
	Knoxville, Tennessee, USA \\
	\texttt{lhan@utk.edu}
    \And
    \href{https://orcid.org/0000-0003-2769-7808}{\includegraphics[scale=0.06]{orcid.pdf}\hspace{1mm}Candace Brakewood} \\
	Department of Civil and Environmental Engineering\\
	University of Tennessee, Knoxville\\
	Knoxville, Tennessee, USA \\
	\texttt{cbrakewo@utk.edu}
}

\renewcommand{\headeright}{}
\renewcommand{\undertitle}{}
\renewcommand{\shorttitle}{GTFS2STN: Analyzing GTFS Transit Data By Generating Spatiotemporal Transit Network}

\hypersetup{
pdftitle={GTFS2STN: Analyzing GTFS Transit Data by Generating Spatiotemporal Transit Network},
pdfsubject={q-bio.NC, q-bio.QM},
pdfauthor={Diyi ~Liu, Jing ~Guo, Yangsong ~Gu, Meredith ~King, Lee D. ~Han, Candace ~Brakewood},
pdfkeywords={Transit System, General Transit Feed Specialization, Transit Assessibility, Travel Time Variability},
}

\begin{document}
\maketitle

\begin{abstract}
	The General Transit Feed Specification (GTFS) is an open standard format for recording transit information, utilized by thousands of transit agencies worldwide. This study introduces GTFS2STN, a novel tool that converts static GTFS transit networks into spatiotemporal networks, connecting bus stops across space and time. This transformation enables comprehensive analysis of transit system accessibility. Additionally, we present a web-based application version of the GTFS2STN tool that allows users to generate spatiotemporal networks online and perform basic analyses, including the creation of isochrone maps from a given origin and the calculation of travel time variability between origin-destination pairs over time. Comparative analysis demonstrates that GTFS2STN produces results similar to those of Mapnificent, an existing open-source tool for generating isochrone maps from GTFS inputs. Compared with Mapnificent, GTFS2STN offers enhanced flexibility for researchers and planners to evaluate transit plans, as it allows users to upload and analyze historical or suggested GTFS feeds from any transit agency. This feature facilitates the assessment of accessibility and travel time variability in transit networks over extended periods, making GTFS2STN a valuable tool for the planning and research for the transit systems.
\end{abstract}


\section{Introduction}
The General Transit Feed Specification (GTFS), initially proposed by Google as the Google Transit Feed Specification, has been widely adopted by transit agencies since the early 2010s to share transit schedules with the public via the internet (\cite{GTFS}). With historical GTFS feeds archived on platforms such as \cite{transitland}, GTFS has emerged as a valuable research resource for transit analysis. This historical data enables researchers to compare past versions with current feeds, providing insights into how transit agencies have evolved their services over time.

\begin{figure}
    \centering
    \includegraphics{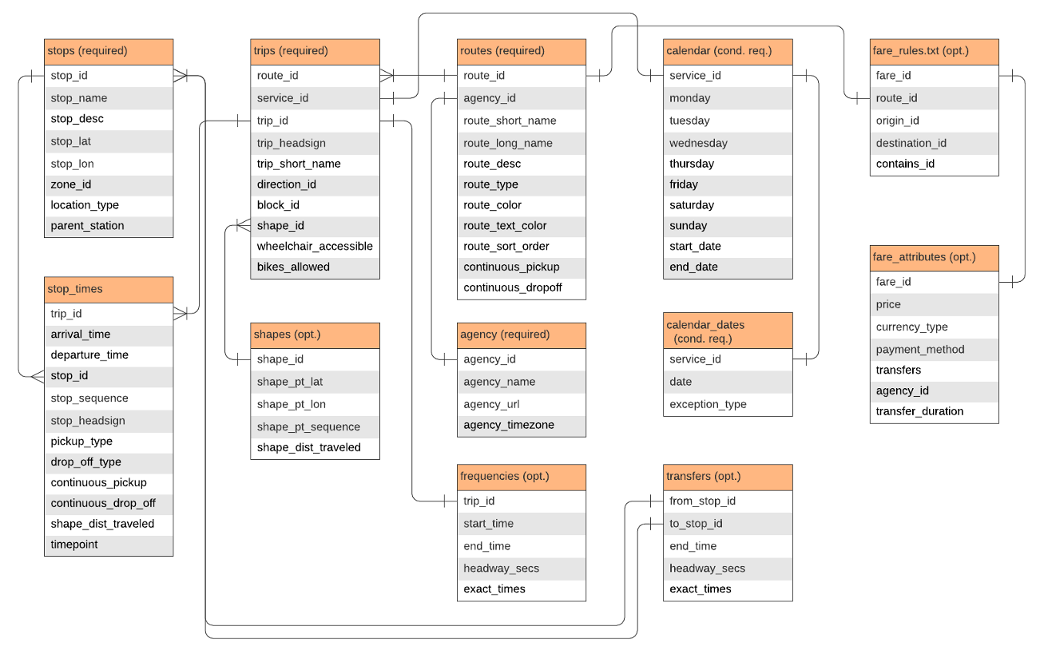}
    \caption{The Entity Relationship Diagram (ERD) of the GTFS data set}
    \label{fig:erd}
\end{figure}

Each GTFS feed is represented using multiple tables as a data set, which encapsulates the complete transit service of an agency for a specific date range. The feed consists of several mandatory and optional tables, structured similarly to a typical SQL database with primary and foreign keys. Figure \ref{fig:erd} illustrates the relationships between these tables through an Entity Relationship Diagram (ERD). This diagram encompasses both required tables (e.g., agency, trips, stops, routes) and optional ones (e.g., shapes, frequencies, transfers, fare rules and attributes). Each table represents a distinct aspect of the transit system, with primary and foreign keys establishing connections between different tables. Collectively, these interconnected tables construct a comprehensive static representation of the entire transit system.

Understanding the GTFS data structure facilitates the development of various applications, including route planning (e.g., \cite{ORS}, \cite{Mapnificent}), public service information provision (e.g., Google Maps, Open Trip Planner), system visualization and analysis, etc. Nearly all these applications rely on the construction of a spatiotemporal network. To address this fundamental need, we propose GTFS2STN, a standardized tool designed to generate spatiotemporal transit networks as the foundation for comprehensive transit analysis.

The remainder of this paper is structured as follows: Section \ref{sec:lit_review} reviews previous GTFS-based studies, focusing on three key aspects: transit accessibility, transit data visualization, and travel time variability. Section \ref{sec:method} details the process of constructing spatiotemporal networks and outlines the algorithms used to generate outcomes. Section \ref{sec:experiment} presents case studies and demonstrates the basic functionalities of the GTFS2STN application.

\section{Literature Review}
\label{sec:lit_review}

Since the advent of GTFS, numerous studies have utilized this data for research purposes. As highlighted by \cite{antrim2013many}, GTFS has proven valuable for various applications, including trip planning, data visualization, accessibility analysis, and urban planning. Among these applications, transit accessibility/variability and transit data visualization are two major research topics particularly relevant to this study, which we will explore in detail in this section.

\subsection{Transit Accessibility}
One of the most extensively studied topics is the measurement of transit system accessibility across time and space using GTFS data. For instance, \cite{farber2014temporal} examined the temporal variability of transit-based accessibility to supermarkets. Their study analyzed how accessibility from census blocks to the nearest supermarkets fluctuates over time. They identified "food deserts" - areas lacking adequate access through public transport and within walking distance - by considering variations throughout the day as well as mean travel times. Furthermore, by incorporating demographic information from census blocks, the study analyzed gender and racial equity in terms of food access.

While \cite{farber2014temporal} employed shortest path algorithms for their analysis, which is a common approach in accessibility studies, other researchers have explored alternative methodologies. \cite{wessel2017discovering} conducted a comparative analysis between static GTFS data and real-time vehicle location data from NextBus. Their study identified times and locations with a higher likelihood of delays, suggesting areas where schedule padding might be necessary. A later study by \cite{wessel2019accuracy} tried to compare the gap of variability measures between GTFS data and the automatic vehicle location (AVL) data. They used GTFS and AVL data to regenerate the travel network. The variation in actual operation causes some remote places a worse level of accessibility. It is found that travel fluctuations contribute to estimating traveling time. In another approach, \cite{goliszek2016use} combined GTFS data with OpenStreetMap network information to create minute-based isochrones. Recognizing the importance of considering both supply and demand in transit system analysis, \cite{fayyaz2017dynamic} proposed an analytical framework to measure transit accessibility while accounting for temporal fluctuations. Their method incorporates indicators to identify causes of poor accessibility, providing a more comprehensive understanding of transit system performance. \cite{polzin2002development} proposed a framework combining demand and supply of transit system to measure time variability.

In summary, these diverse approaches demonstrate the versatility of GTFS data in transit research and highlight the importance of considering temporal and spatial variations in accessibility studies.

\subsection{Transit Data Visualization \& Analysis}

Visualizing transit systems offers an effective method to further exploit and understand GTFS data. \cite{prommaharaj2020visualizing} explored several techniques for visualizing public transit systems using GTFS data. Six different visualization modules (i.e., mobility, speed, flow, density, headway, and analysis) are introduced. The researchers utilized various diagrams to visualize headway patterns throughout the day. Additionally, they implemented a top list feature to identify extreme data points, such as the busiest stations or those with the longest waiting times. This approach enables transit planners and researchers to quickly identify areas of concern or exceptional performance within the system.

Beyond visualization, GTFS data can be leveraged to analyze more technical metrics of transit systems. \cite{wong2013leveraging} demonstrated how GTFS data could be used to measure the Level of Service (LOS) as defined in the Transit Capacity and Quality of Service Manual (TCQSM) for transit agencies. Their study examined metrics such as average headway, stop spacing, and other relevant indicators. Furthermore, \cite{wong2013leveraging} recognized the need to evaluate LOS separately for different transit modes including bus, light rail, subway, and commuter rail. 

While most studies have processed and visualized GTFS data on local machines, some researchers have endeavored to engineer online or real-time visualization solutions. One of the most ambitious applications in this domain is the real-time transit data visualization system proposed by \cite{bast2014real}. This innovative system creates a worldwide live map that demonstrates real-time information of transit systems across the globe. To achieve this ambitious goal, \cite{bast2014real} employed several sophisticated techniques, including time-space queries, interpolated schedule, and spatial-temporal bounding boxes. These methods are utilized to significantly reduce response times on the client side, ensuring a smooth and responsive user experience despite the vast amount of data being processed. Notably, the system uses GTFS data as a fallback when real-time positional data is unavailable, demonstrating the continued importance of GTFS even in advanced, real-time applications.

Despite the abundance of transit service analyses based on GTFS data, there is a notable absence of an open-source tool specifically designed to generate spatiotemporal networks for transit analysis. The development of such a tool is both necessary and important, as it would provide a standardized method for expanding GTFS data into a spatiotemporal network - essentially creating the skeletal structure for comprehensive transit analysis.

This identified gap in the existing literature and toolset serves as the primary motivation for our current study. By developing a tool that can consistently and efficiently transform GTFS data into spatiotemporal networks, we aim to facilitate more advanced, standardized, and comparable transit analyses across different systems and studies. We anticipate fostering more comprehensive and nuanced understanding of transit systems, ultimately contributing to improved public transportation planning and operations.

\section{Methodology}
\label{sec:method}

The methodology for this study comprises three main components: (1) A basic example of spatiotemporal network; (1) Generation of spatiotemporal network; (2) Path searching algorithms;

\subsection{A Basic Example of Spatiotemporal Network}
\begin{figure}[h]
    \centering
    \includegraphics[width=\textwidth]{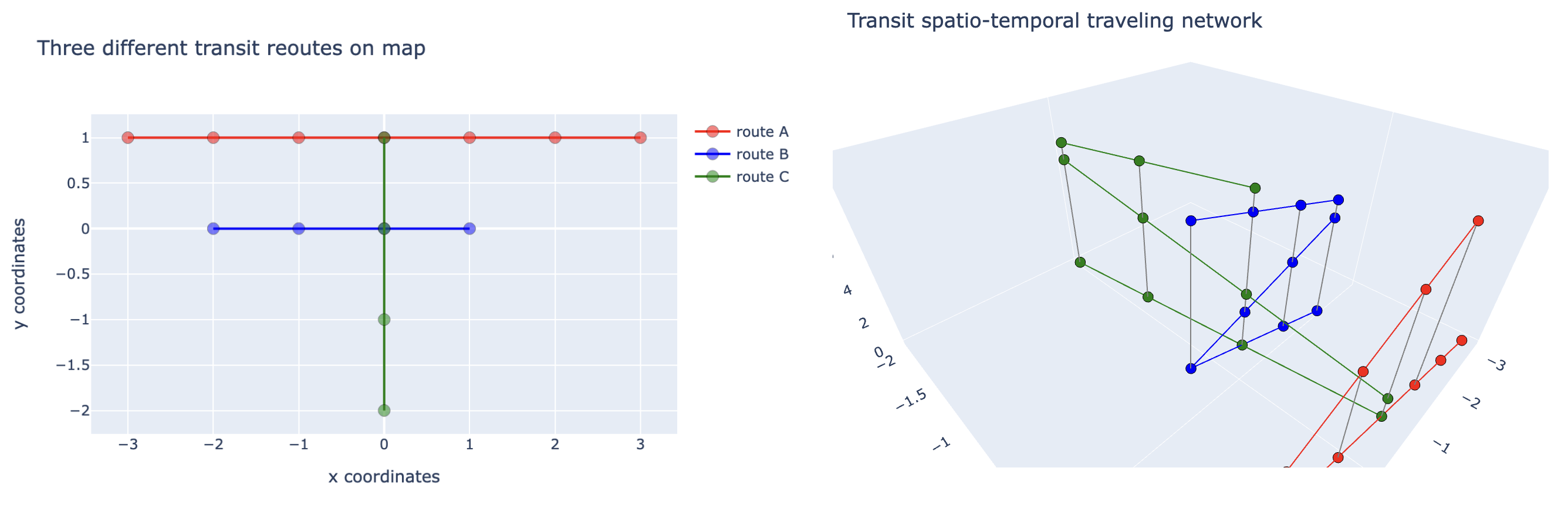}
    \caption{A simple demonstration of converting a transit network to a spatiotemporal transit network}
    \label{fig:build-network}
\end{figure}

Traditional static networks are insufficient for comprehensive travel time analysis, as transit services vary throughout the day. To address this limitation, we expand the network across the time dimension, creating a spatiotemporal network. Figure \ref{fig:build-network} illustrates the process of converting bus routes into a three-dimensional spatiotemporal transit network. The left sub-figure in Figure \ref{fig:build-network} depicts three distinct traffic routes overlaid on a map. The right sub-figure introduces an additional dimension - a time axis representing the time of day. This example showcases three buses traveling back and forth along three routes. Specifically, In the three-dimensional spatiotemporal network (right sub-figure), the vertical lines represent passengers' ability to wait at transit stops over time. Although not explicitly shown in Figure \ref{fig:build-network}, passengers can walk between different transit stops to access other routes. Besides, each node in the network corresponds to a specific location at a particular time. This comprehensive approach allows for a more nuanced and realistic representation of transit systems, capturing the dynamic nature of public transit schedules throughout the day.

\subsection{Generate Spatiotemporal Network}
\begin{figure}[h]
    \centering
    \includegraphics[width=\linewidth]{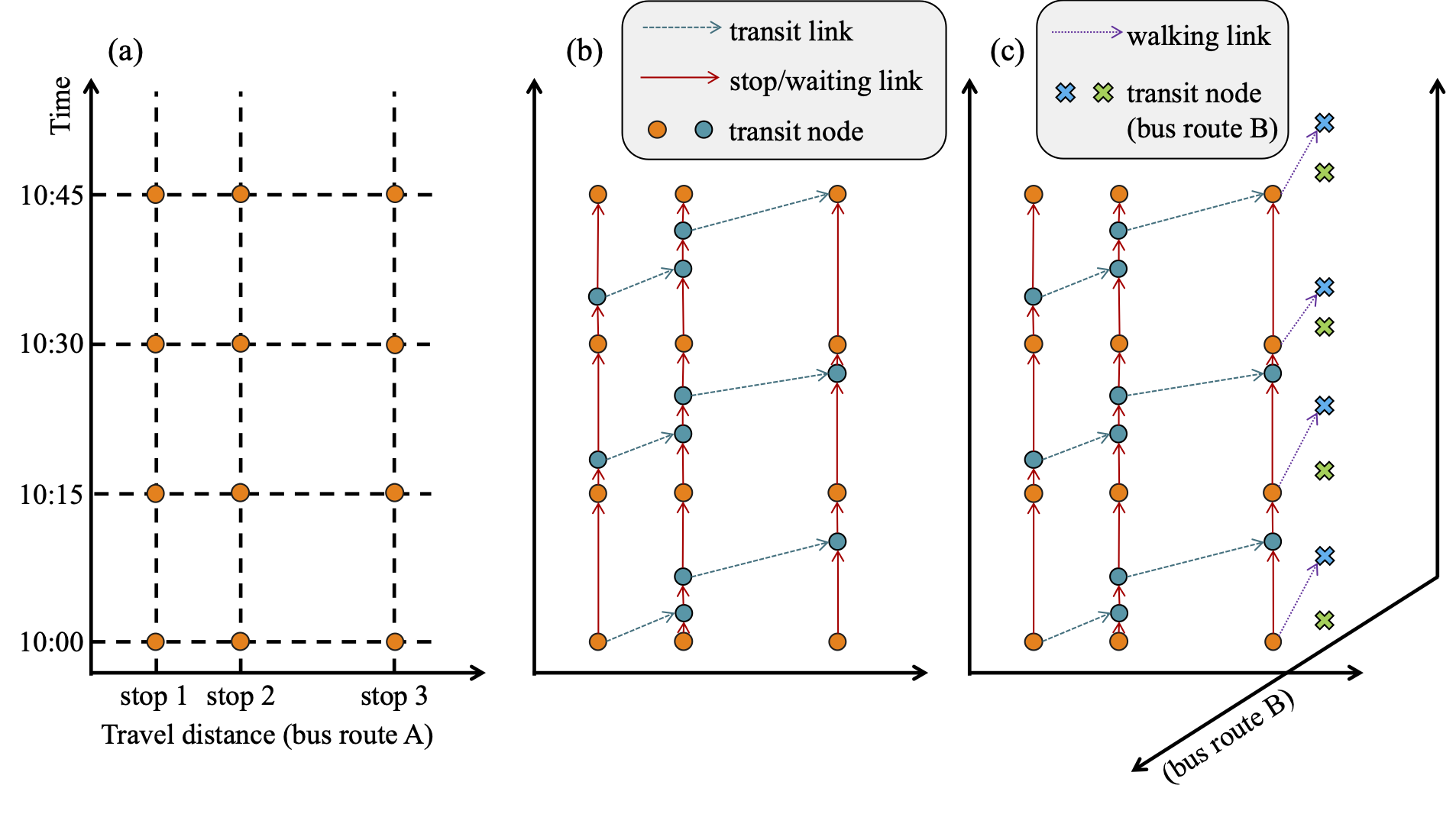}
    \caption{A simplified example of generating spatiotemporal network for three consecutive stops of a transit route}
    \label{fig:build_gtfs_network}
\end{figure}

The process of generating the spatiotemporal traffic network involves several interconnected steps. The first step is to create duplicated stop nodes across the time dimension, effectively representing each stop at various times throughout the day. This forms the foundation of our temporal expansion. For example, Figure \ref{fig:build_gtfs_network} (a) shows the skeleton network of a segment of a bus route. The x-axis shows the distances of consecutive stations, whereas the y-axis shows the time dimension. One bus stop node represents the status of one node at a given time.

After generating the skeleton nodes, the second step adds more bus nodes and links based on the bus time schedule, as represented in Figure \ref{fig:build_gtfs_network} (b). The blue links with dotted lines represent a transit vehicle traveling from one stop to another. Besides the skeleton nodes in Figure \ref{fig:build_gtfs_network} (a), more nodes are added based on the transit time schedule.

Besides traveling buses, one needs to consider walking distance to access bus stops or transfer between close bus stops. To account for pedestrian movement between stops, we define a maximum walking distance buffer (e.g., 0.25 miles). For each bus stop node, we then add walking edges to all neighboring nodes that fall within this buffer distance. For example, the purple links in Figure \ref{fig:build_gtfs_network} (c) shows the transfer links from skeleton bus stop nodes to another bus route.

Finally, after adding all the transit traveling links and walking links, we can finally connect all the links representing the same bus stop in time order along the time direction. These links are called stop or waiting links representing the possibility of a traveler waiting at a transit stop. As a demonstration, the red solid lines in \ref{fig:build_gtfs_network} (b)-(c) represents the stop/waiting links.

In summary, the waiting links, the transit links, and the walking links togehter formed all the links of the transit travel network. Combined with the stop nodes at both ends of the network's links, a transit travel network is fully generated. Table \ref{tab:network_elements} summarizing the four components of the basic spatiotemporal network, making it possible to query the shortest path between any two stop nodes at specific times.

\begin{table}[h]
    \centering
    \caption{Four basic elements of spatiotemporal network}
    \begin{tabular}{c c}
        \hline
        Name & Description \\
        \hline
        stop node & traffic nodes of network (i.e., a bus stop at a give time) \\
        stop/waiting link & vertical links connecting the same stop over time \\
        transit link & links connecting different bus stops traversed by buses \\
        walking link & links connecting different bus stops traversed by walking \\
        \hline
    \end{tabular}
    \label{tab:network_elements}
\end{table}

To facilitate more comprehensive analysis, we introduce an additional layer of abstraction. For each transit stop, we generate a single origin node and a single destination node. These nodes connect to all temporal instances of their respective stop. This enhancement allows for more flexible querying. For instance, analysts can request the shortest path to a specific stop without specifying an arrival time because all links are pointed towards the corresponding destination node of the link.

\begin{figure}[h]
    \centering
    \includegraphics[width=.8\textwidth]{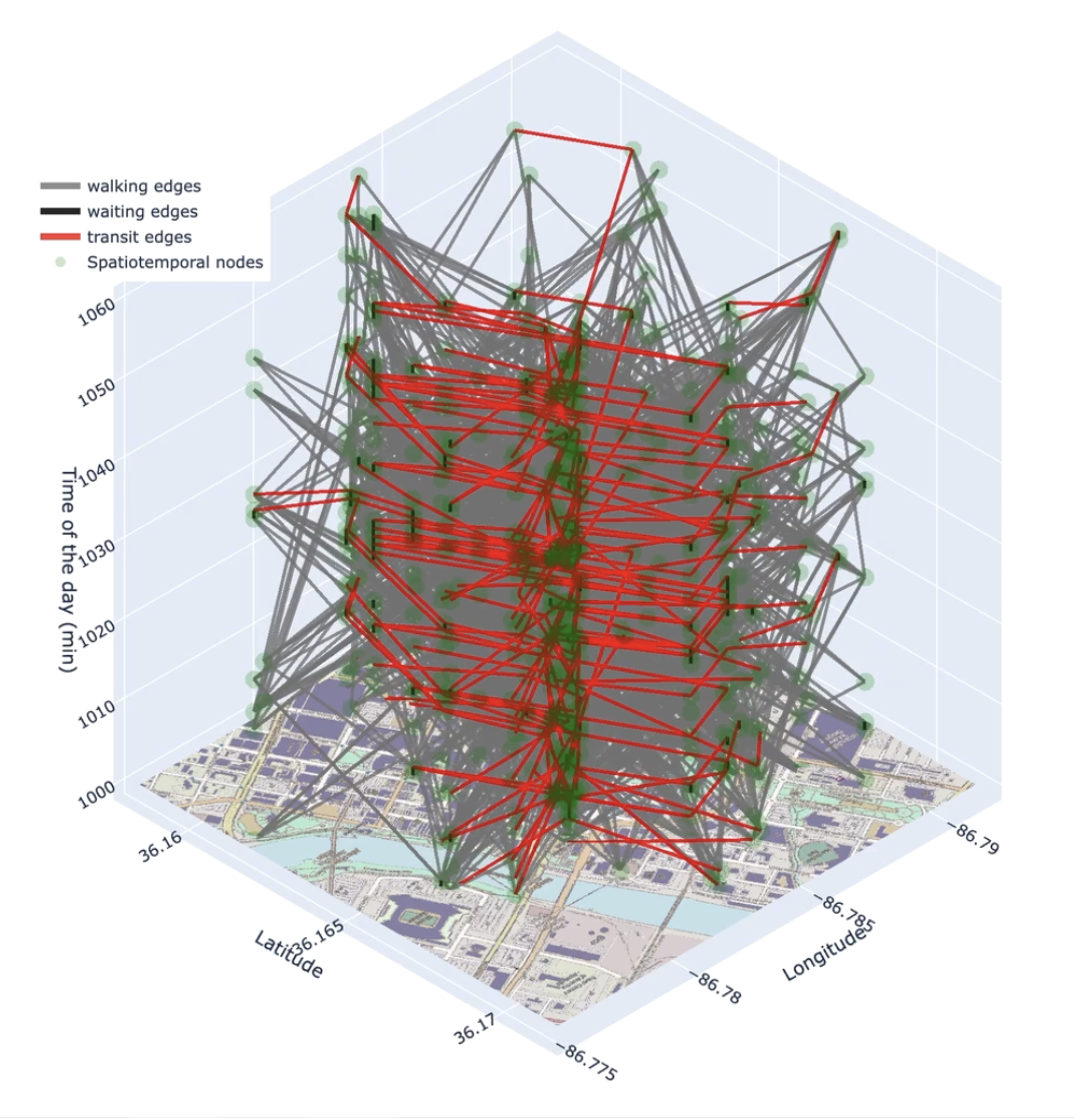}
    \caption{a spatiotemporal network generated in downtown Nashville, TN (a small segment of the network)}
    \label{fig:network-nashville}
\end{figure}

Note that the steps in Figure \ref{fig:build_gtfs_network} mainly focuses on three consecutive bus stops along a transit route. A more comprehensive example is provided in Figure \ref{fig:network-nashville} by visualizing a small segment of the spatiotemporal network in downtown Nashville, Tennessee. The bottom map shows the corresponding topologies by longitudes and latitudes. The vertical dimension shows the time of the day. The red and grey links are traversed by transit vehicles and by walking, respectively. The black links are the stop/waiting links. Each green dot represents the node of a bus stop as a given time. Building such a network can help query the shortest travel time between two places.

\subsection{Path Searching Algorithms}
Once the spatiotemporal network is generated, it becomes possible to search for travel times between different locations using shortest path searching algorithms. In this study, we employ Dijkstra's algorithm for path searching due to its efficiency and reliability in finding the shortest path in a weighted graph.

\begin{figure}[h]
    \centering
    \includegraphics[width=.8\textwidth]{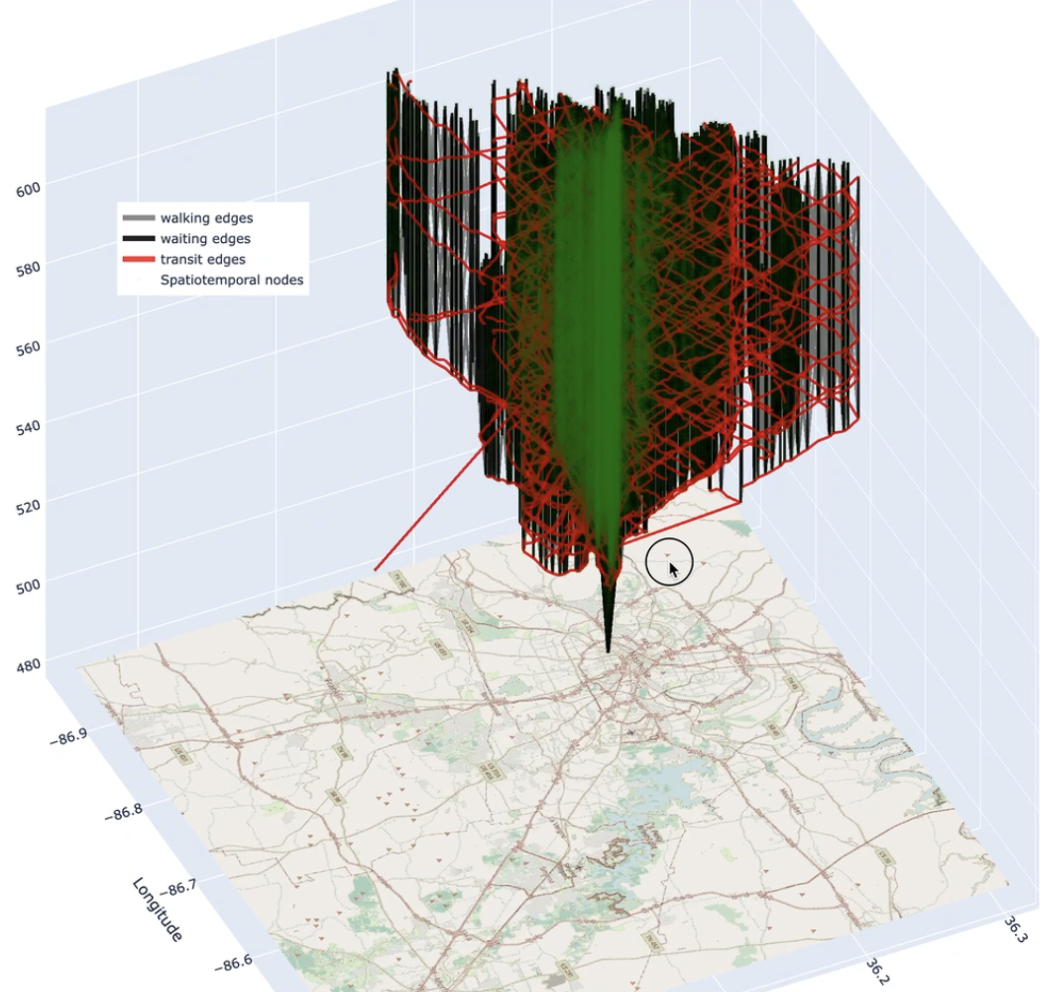}
    \caption{A shortest paths starting from a bus stop in Nashville, TN}
    \label{fig:network-nashville-path}
\end{figure}

The flexibility of our approach allows for searching shortest paths given either a set of origins or a set of destinations. For example, Figure \ref{fig:network-nashville-path} illustrates this concept by displaying the sub-network that can be traversed at different times of the day from a given origin using in Nashville, Tennessee.

The generated spatiotemporal diagram offers significant potential for addressing various transit-related problems. By modifying the network's topology, many flexible queries become available. For instance, by reversing all the links in the network, we can generate isochrone plots to specific destinations, providing insights into inbound accessibility. Furthermore, by introducing a hyper destination node connected to several destination stop nodes, we can analyze isochrone plots for multiple origins or destinations simultaneously. This approach proves particularly useful when studying accessibility to a group of locations, such as healthcare facilities or employment centers. Finally, it is possible to query the shortest traveling time between several origins and several destinations by adding hyper-nodes.

In summary, the method of generating spatiotemporal diagrams is crucial for analysis. The applications of the method can extend beyond the aforementioned examples. For instance, the spatiotemporal network can be adapted to analyze temporal variations in service frequency, identify optimal transfer points, or evaluate the impact of service disruptions on overall network accessibility. By providing a comprehensive framework for representing both the spatial and temporal aspects of transit systems, our approach opens up new avenues for in-depth analysis and optimization of public transportation networks.

\subsection{More visualization approaches}
After searching for the shortest paths between origins and/or destinations, more visualization options are available for interpreting the results besides isochrone map. If both origins and destinations are known, the shortest traveling time can be viewed as a diagram with respect to time of the day. The journey/travel time can be decomposed to three parts, including the walking time (in blue), the waiting time (in orange), and the time spend on buses (in orange).

Another visualization method is to analyze the service frequency at each locations. One way of visualizing it is by splitting the region into small squares using grids. For example, a dense grids split the service region of the WeGo transit in Nashville is visualized in Figure \ref{fig:tt} below. No matter how the layout of the transit system changes, information can be gathered spatially using the array of grids. By counting the visiting frequencies within the grid, one can aggregate the average visiting density within each small square to understand the service level.

\begin{figure}
    \centering
    \includegraphics[width=0.8\linewidth]{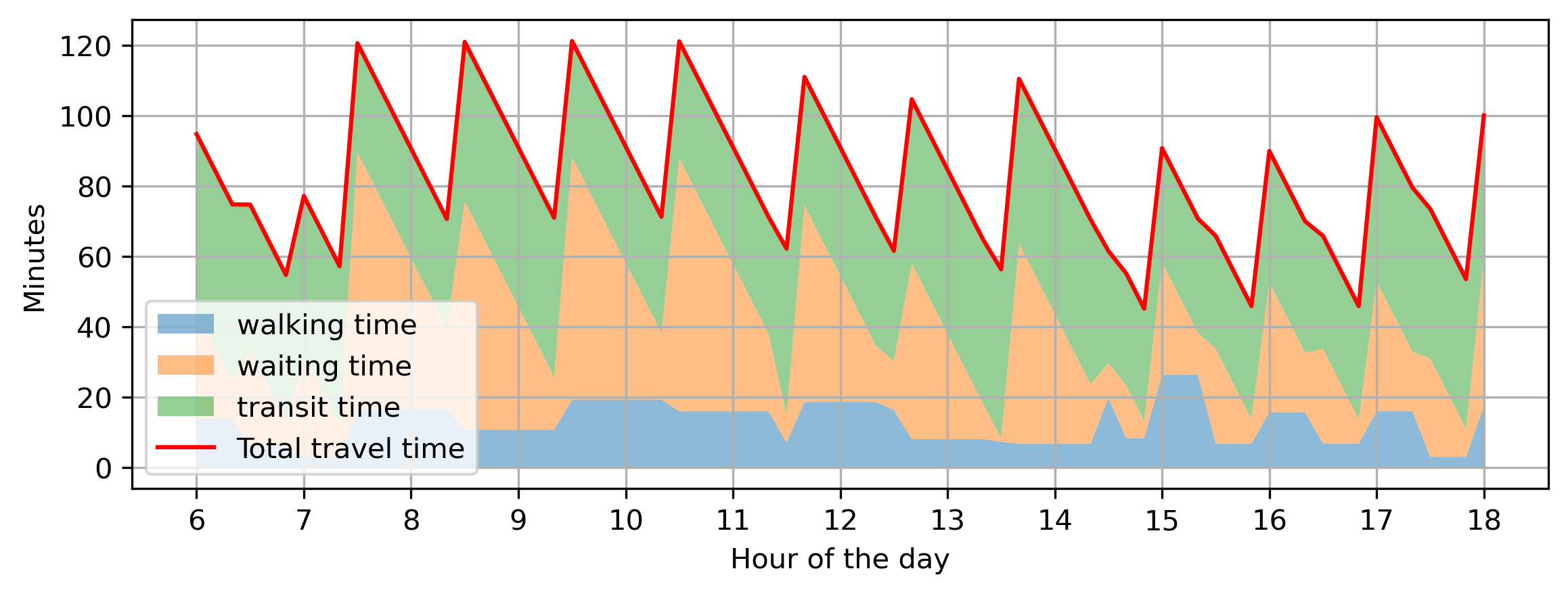}
    \caption{The journey time between an origin-destination pair over day time}
    \label{fig:tt}
\end{figure}

\begin{figure}[h]
    \centering
    \includegraphics[width=\linewidth]{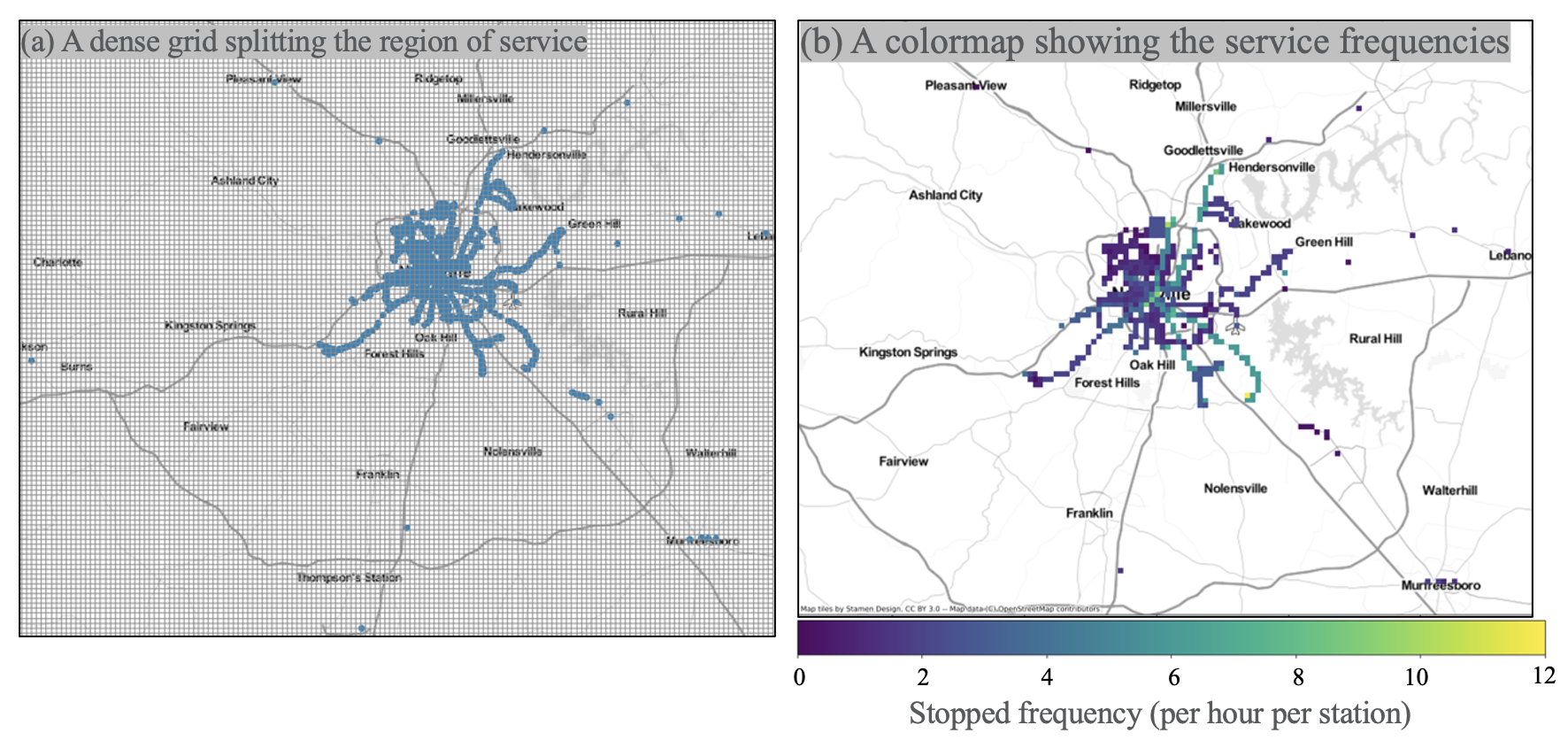}
    \caption{Analyzing service frequencies at different locations by aggregating over a dense grid}
    \label{fig:grids}
\end{figure}

Figure \ref{fig:grids} shows the stop’s visiting frequency using the grid method.  In Figure \ref{fig:grids} (b), the colored grids are the places with at least one stop within the grid. Since the average is taken within each grid, the number of visits in the busiest grid zone is only 12 rather than over 20s for outlying transit stops. The following Equation \ref{eq1} is used to calculate the averaged frequency of stop visits within the region.

\begin{equation} \label{eq1}
    \mbox{Averaged frequency of stop visits} = \frac{\mbox{Number of stops within the region}}{\mbox{Total number of stops} \times \mbox{Number of observed hours}}
\end{equation}

\section{Examples \& Usages}
\label{sec:experiment}
To demonstrate the versatility and applicability of our GTFS2STN tool, we conducted a series of experiments using the WeGo transit system in Nashville, Tennessee. While the tool is designed to work with any standard GTFS data feed, we chose to focus on a single transit system for consistency throughout this paper. Our analysis explores various scenarios to gain insights into the system's performance and accessibility. To validate our results, we compared them with outputs from Mapnificent, another tool that generates real-time isochrone maps using GTFS data inputs.

\subsection{A Step-by-step Guide of Using the Application Version of the GTFS2STN}

\begin{figure}
    \centering
    \includegraphics[width=\textwidth]{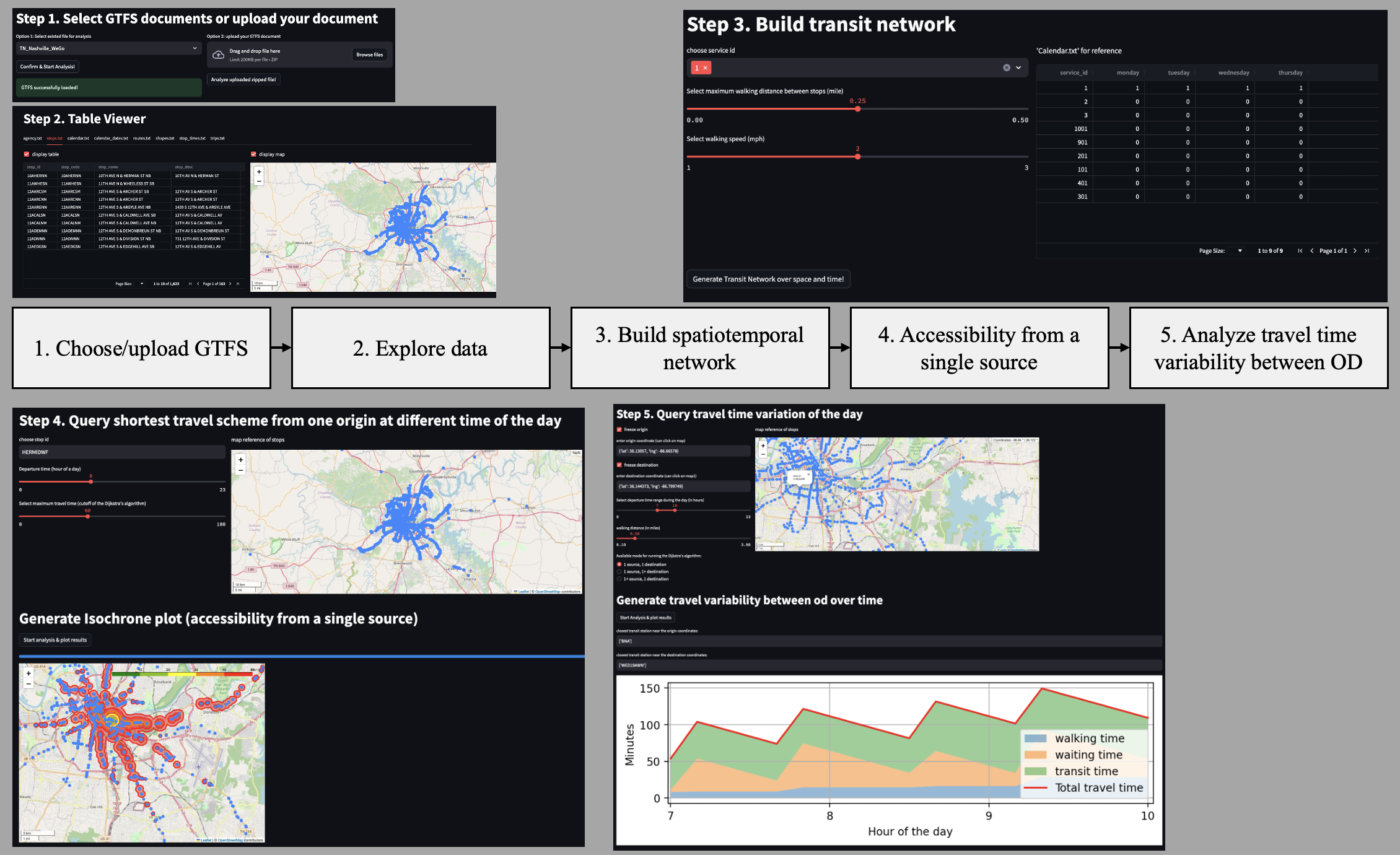}
    \caption{The 5 major steps of the using the GTFS2STN application}
    \label{fig:gtfs2stn_steps}
\end{figure}

The GTFS2STN application comprises five major steps for network analysis, as illustrated in Figure \ref{fig:gtfs2stn_steps}. The process begins with data input. Users can either select an existing file or upload their own GTFS document. After loading the data, users confirm their selection to proceed to the next stage.

In the second step, users can visualize each table in the dataset. For tables containing geographical information, such as ``stops.txt'' or ``shapes.txt'', the application offers an interactive map view. Users can explore the system by hovering over and clicking on various points of interest.

The third step involves building the network. Users select one or multiple service IDs from the ``calendar.txt'' file to define the operating scope. Additionally, they specify the maximum allowable walking distance and speed to establish walking links. The spatiotemporal network is then generated, and users have the option to download it for further analysis.

The fourth step focuses on accessibility analysis. Users select an origin transit stop, departure time, and maximum allowable journey time (cutoff time). Based on these parameters, the application generates isochrone maps to visualize accessibility.

The final step analyzes travel times between specified origins and destinations. Users can either click on the map to select stops or manually input coordinates. Upon initiating the analysis, the application visualizes the journey time, breaking it down into three components: walking time (blue), waiting time (orange), and transit time (green). A red line at the top represents the total journey time, summing up these three segments.

This step-by-step approach allows for a comprehensive analysis of the transit system, providing insights into accessibility, travel times, and network efficiency. By offering both visual and quantitative outputs, GTFS2STN enables users to gain a nuanced understanding of the transit system's performance across various scenarios and parameters.

\subsection{Case Study: Accessibility to Walmart markets in Nashville, Tennessee}
The flexibility of our GTFS2STN tool, as outlined in the methodology section, allows for diverse and insightful analyses through the addition of hyper nodes and links. To illustrate this capability, we conducted an accessibility analysis focusing on Walmart markets in Nashville, Tennessee.

\begin{figure}[h]
    \centering
    \includegraphics[width=\textwidth]{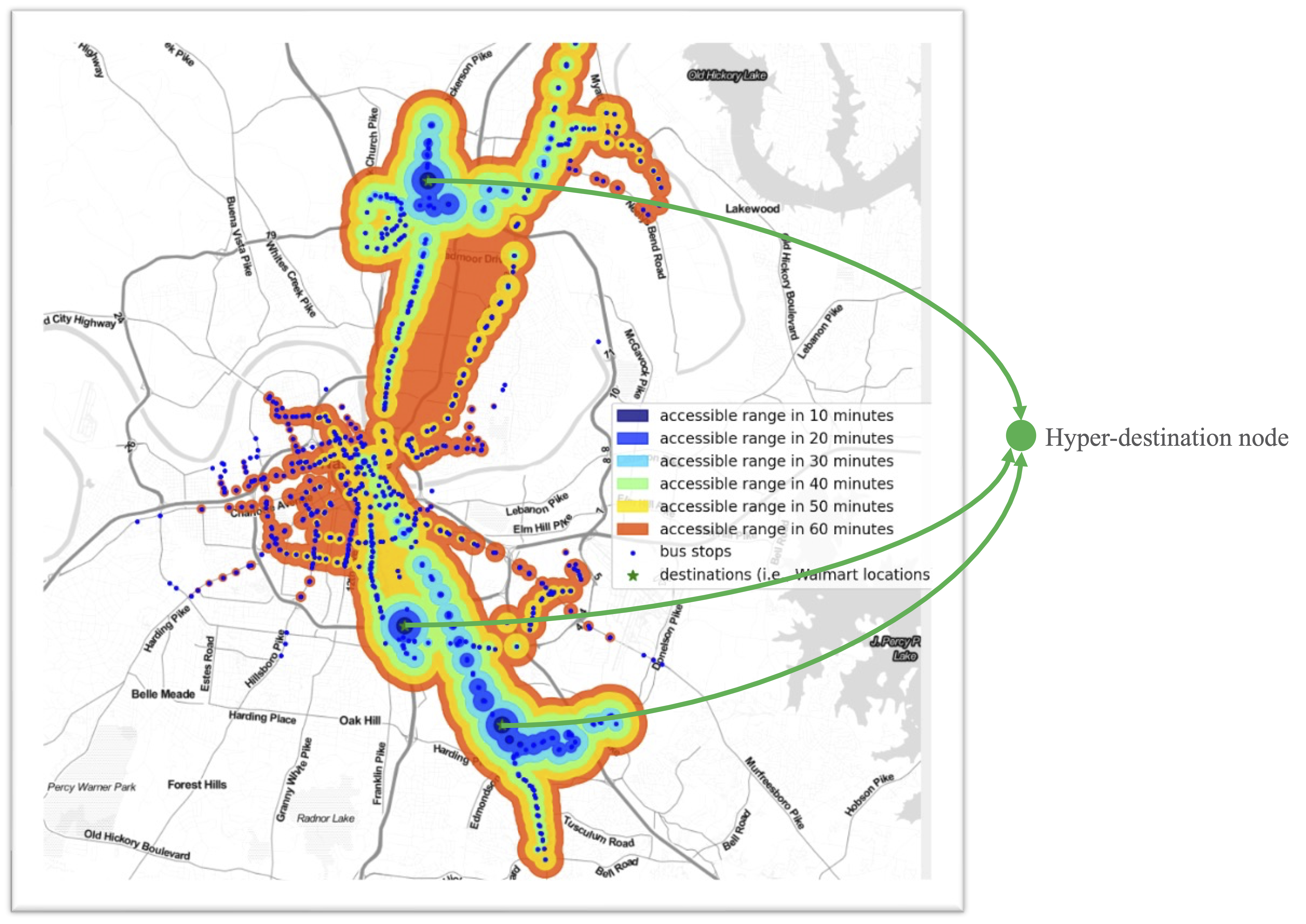}
    \caption{The isochrone map to access any of the three Walmart markets in Nashville, Tennessee}
    \label{fig:walmart}
\end{figure}

Figure \ref{fig:walmart} identifies three Walmart locations within the city. We generated an isochrone plot with a 60-minute travel time threshold to visualize the accessibility of these markets via public transit. The results reveal that these Walmart locations, collectively, are accessible to a substantial portion of the city, primarily along major arterial routes. This analysis demonstrates the tool's ability to assess accessibility to multiple destinations simultaneously, which can be particularly valuable for urban planning and retail strategy development.

\subsection{Case Study: Temporal Variations in Accessibility}
\begin{figure}[h]
    \centering
    \includegraphics[width=\textwidth]{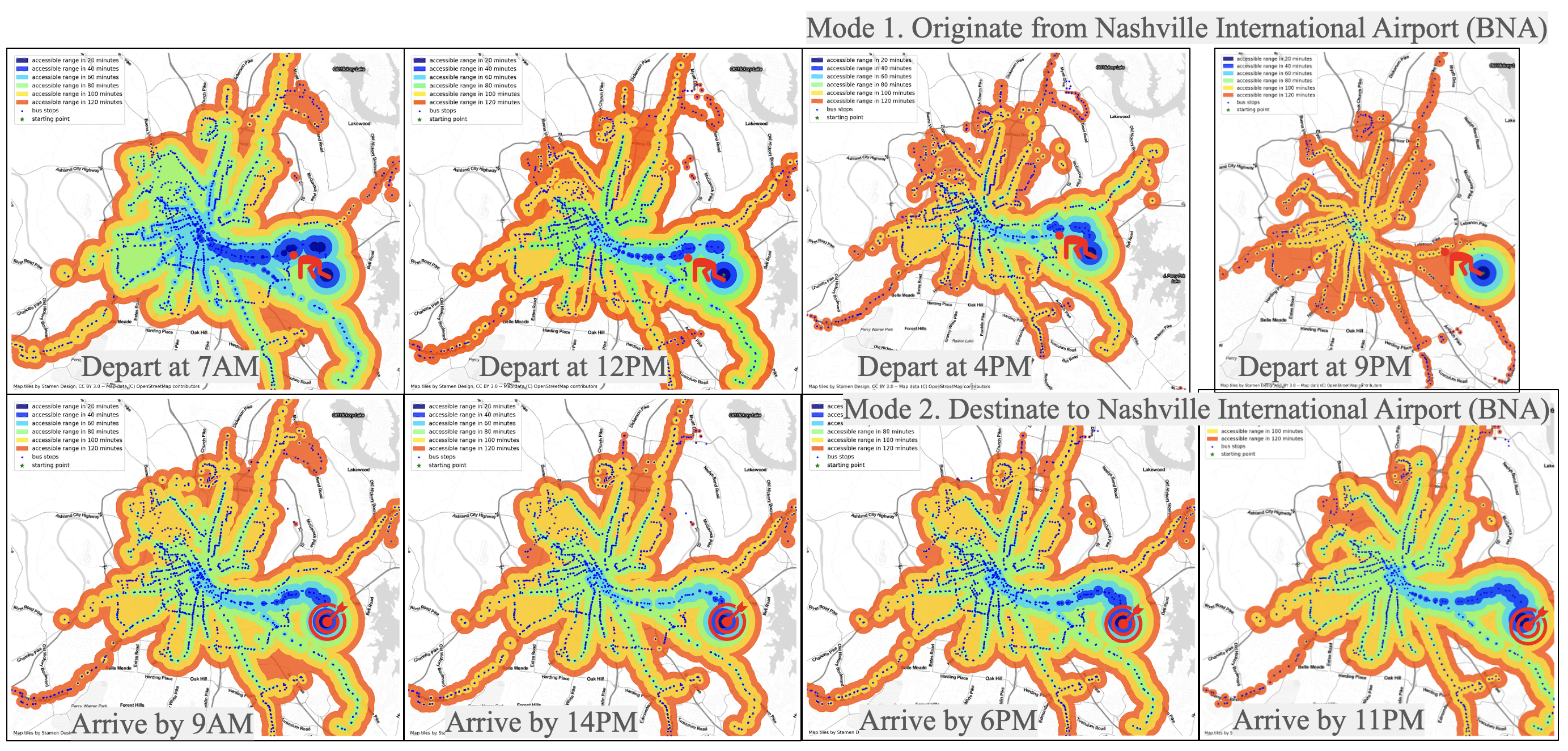}
    \caption{Accessibility from/to the Nashville International Airport (BNA) using WeGo transit services in Nashville, Tennessee}
    \label{fig:bna}
\end{figure}

To further showcase the tool's capabilities in capturing temporal dynamics of transit systems, we conducted a case study examining accessibility levels at different times of the day. For this analysis, we focused on trips originating from and terminating at the Nashville International Airport, as shown in Figure \ref{fig:bna}. Each subplot is an isochrone plot. The first row of subplots shows the accessibility originating from Nashville International Airport (BNA) airport starting from different time of the day, whereas the second row of subplots shows the accessibility destination to the BNA airport with different latest arrival time.

By comparing different isochrone plots of in Figure \ref{fig:bna}, the study reveals significant variations in service levels throughout the day. Most notably, we observed that transit accessibility is considerably limited at 9 PM. This stark contrast in service availability highlights the importance of considering temporal factors in transit planning and analysis. Such temporal accessibility analyses can provide crucial insights for various stakeholders: (1) Transit planners can identify periods of limited service, informing decisions about route modifications or service frequency adjustments; (2) Airport authorities can better understand how public transit availability might affect passenger experiences at different arrival or departure times; (3) City officials can assess the airport's connectivity to the broader urban area across different times, which may influence economic development strategies.

\subsection{Comparative Analysis: GTFS2STN vs. Mapnificant}

To validate the effectiveness of our GTFS2STN tool, we conducted a comparative analysis with similar existing tools, particularly focusing on Mapnificent. This comparison provides insights into the accuracy and unique features of our proposed tool.

\begin{figure}
    \centering
    \includegraphics[width=\textwidth]{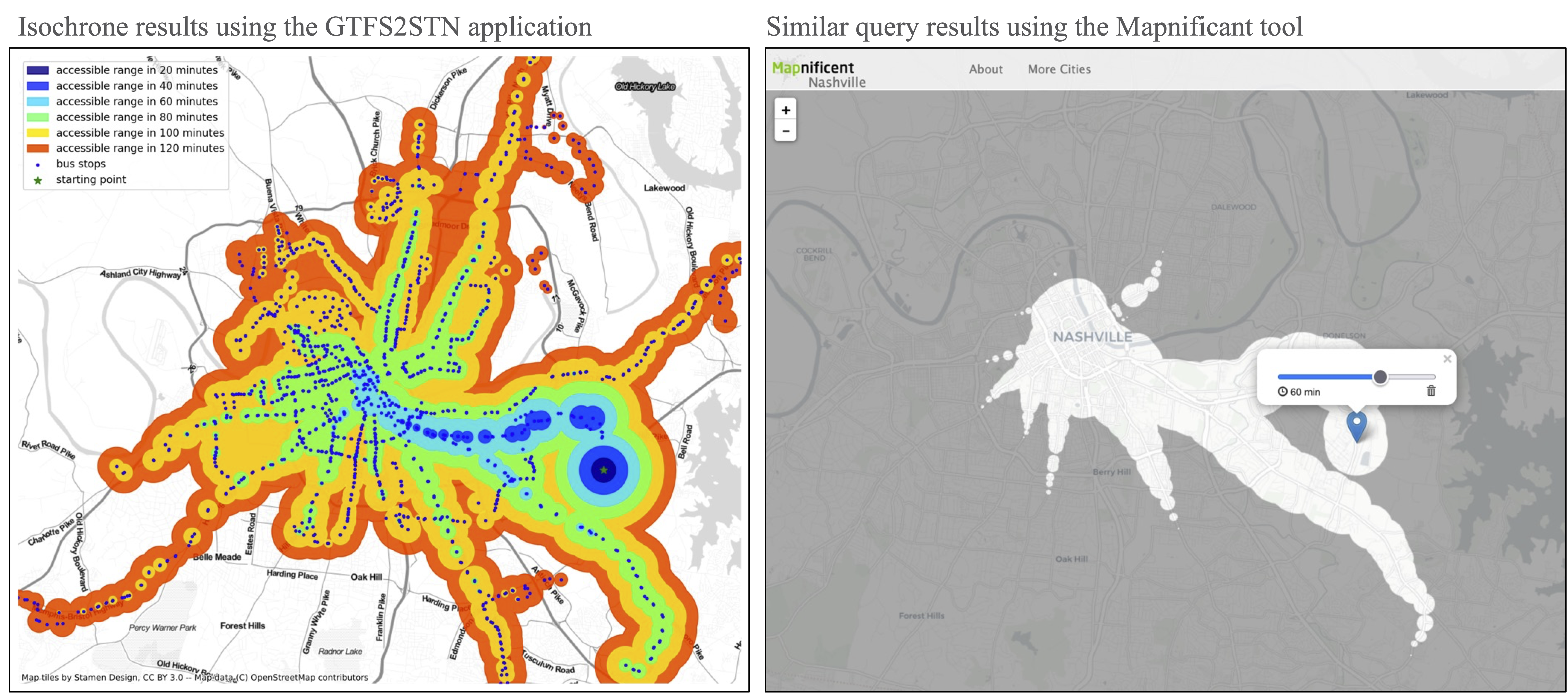}
    \caption{A comparison of isochrone plot between GTFS2STN and Mapnificent using similar query conditions}
    \label{fig:compare}
\end{figure}

Figure \ref{fig:compare} illustrates the results of a query originating from the airport, generated by both GTFS2STN and Mapnificent. While the overall patterns of accessibility are similar, there are notable differences in the presentation and depth of information provided by each tool.

Mapnificent's query is constrained to a 60-minute time bound, presenting a single isochrone boundary. In contrast, GTFS2STN offers a more granular visualization, displaying isochrones ranging from 20 to 120 minutes using a color gradient. This extended range and detailed breakdown allow for a more comprehensive understanding of transit accessibility at various time thresholds.

Upon close examination, we observe that the isochrone generated by GTFS2STN appears slightly smaller than that of Mapnificent. This discrepancy can be attributed to GTFS2STN's more realistic modeling of bus waiting times. By incorporating this additional factor, our tool provides a more conservative, yet potentially more accurate, representation of transit accessibility. Despite this minor difference, the overall accessibility patterns revealed by both tools are remarkably similar. This consistency across different methodologies lends credibility to our results and suggests that GTFS2STN is performing in line with established tools in the field.

The comparative analysis highlights several key strengths of GTFS2STN: (1) Enhanced temporal resolution, allowing for more nuanced accessibility analysis; (2) More realistic modeling of transit experiences by including waiting times; (3) Flexibility in visualizing a wider range of travel times, enabling both broad overview and detailed examination of accessibility patterns.

\subsection{Analyzing the Impacts of Pandemic on the Schedule of the Same Transit System}

To further demonstrate the analytical capabilities of our tool, we conducted a comparative case study using data from the WeGo transit agency in Nashville, Tennessee. This study aimed to contrast transit patterns before and during the Covid-19 pandemic, offering insights into how the pandemic affected public transportation services and usage. This case study explores schedules of two distinct periods in different years, including: (1) April 11th to October 2nd, 2021 (during the pandemic); (2) April 7th to October 5th, 2019 (pre-pandemic baseline)

\begin{figure}
    \centering
    \includegraphics[width=\linewidth]{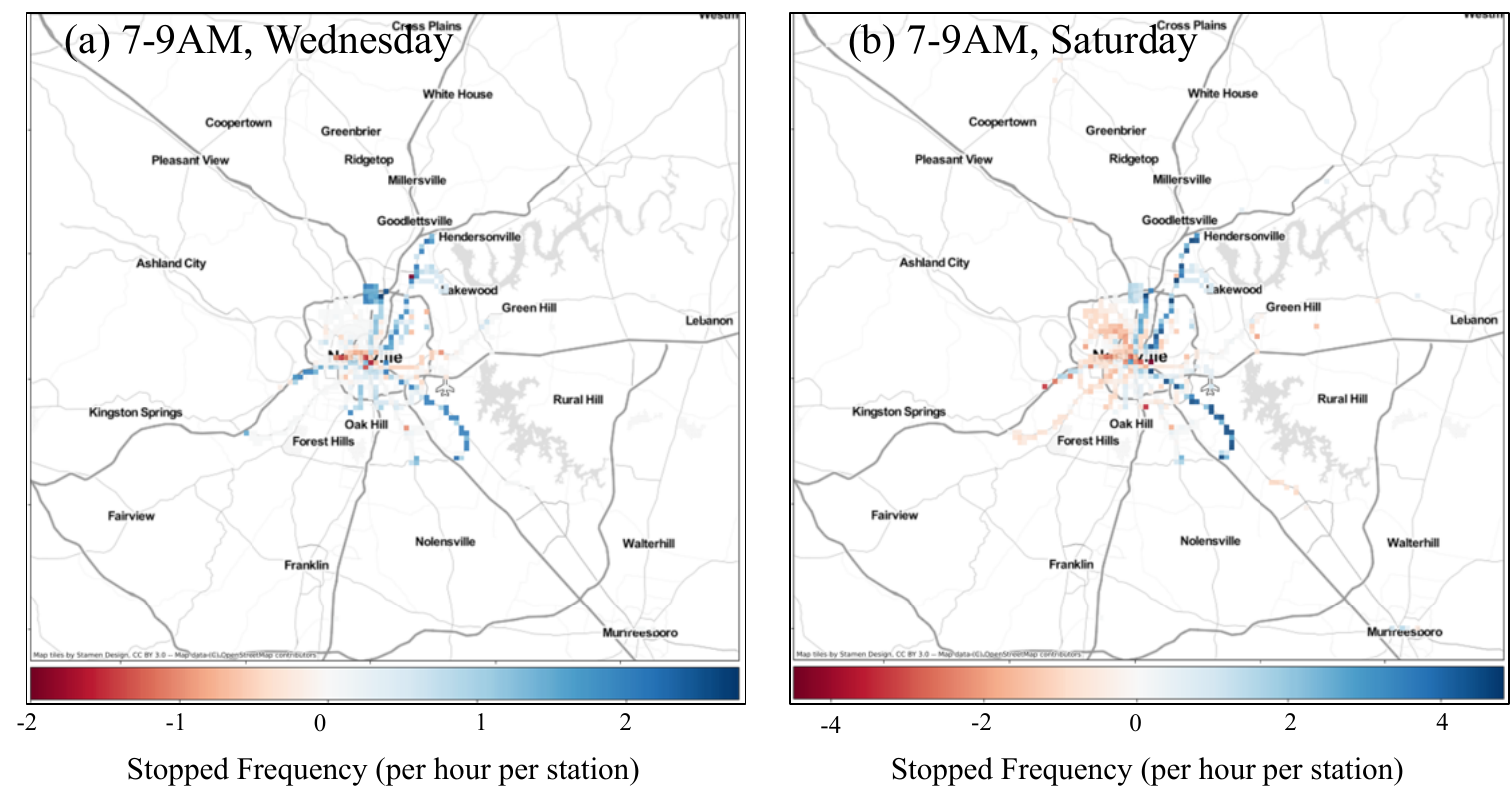}
    \caption{The differences in stopped frequencies of buses (number of stopped buses per hour) between 2021 and 2019}
    \label{fig:transit_compare}
\end{figure}

Figure \ref{fig:transit_compare} illustrates the differences between the 2021 and 2019 scenarios using a diverging color map. This visualization technique employs blue to indicate higher 2021 values, red for higher 2019 values, and white to show areas of no change. This color scheme allows for intuitive interpretation of service changes over time.

Figure \ref{fig:transit_compare} (a) depicts a typical Wednesday morning peak (7-9 AM) comparison. The visualization reveals that North-South corridors (shown in dark blue) experienced more frequent service during weekday peak hours in 2021 compared to 2019. Conversely, the city center and East-West routes saw a slight reduction in service frequency. While not explicitly shown in the figure, our analysis indicates that afternoon peak traffic patterns closely mirror those of the morning, suggesting consistent service frequency throughout weekdays.

Figure \ref{fig:transit_compare} (b) presents the Saturday morning peak (7-9 AM) comparison. Here, we observe a distinct shift in service patterns. The city center experienced reduced service levels, while radial corridor routes saw increased frequency. These findings align with WeGo transit's declared modifications under the Nashville BetterBus project, which aimed to allocate more resources towards longer-distance corridor traffic at the expense of city center coverage.

These results highlight several key points. First, the pandemic prompted a redistribution of transit resources, favoring major corridors over dense urban areas. Second, weekday service patterns remained relatively consistent throughout the day, suggesting a focus on serving commuter needs. Third, weekend service saw more dramatic changes, potentially reflecting shifts in leisure and non-work travel patterns during the pandemic.

In short, this case study demonstrates the power of our tool in visualizing and analyzing complex transit data across different time periods. By enabling such detailed comparisons, the tool provides valuable insights for transit planners and policymakers, helping them understand the impacts of major events like the Covid-19 pandemic on public transportation systems and informing future service adjustments.

\section{Discussion}
\label{sec:discussion}
This study introduces GTFS2STN, a novel application designed to visualize transit accessibility across both time and space. By allowing users to upload any GTFS feed, the tool offers remarkable flexibility for researchers to evaluate historical or projected transit scenarios in terms of accessibility and travel time variability.

GTFS2STN provides a user-friendly interface for interactive data exploration, isochrone plot generation from fixed locations, and journey time analysis between origin-destination pairs. Furthermore, the application enables users to download the generated spatiotemporal transit network, facilitating more in-depth analyses beyond the tool's built-in capabilities.

While our comparative analysis demonstrates that GTFS2STN performs comparably to established tools like Mapnificent, we acknowledge several limitations in its current iteration. Firstly, the walking buffer is currently implemented as a simple circular area, rather than a more realistic network-based buffer that accounts for actual road traversal. Secondly, the tool faces challenges in integrating multiple GTFS feeds simultaneously. This limitation is particularly relevant for large metropolitan areas served by multiple transit agencies, where analyzing a single agency's network may not fully capture the realistic transit scenario. Thirdly, the expansion of the network into a three-dimensional spatiotemporal structure, implemented in Python, can be memory-intensive for large-scale analyses.

To address these limitations and enhance the tool's capabilities, our future research directions include: (1) Implementing a more realistic walking network based on actual road networks; (2) Optimizing memory usage to improve performance for large-scale analyses; (3) Enhancing the tool's accuracy and expanding its analytical capabilities; (4) Developing functionality to integrate multiple GTFS feeds for comprehensive multi-agency analyses.

Despite these current limitations, GTFS2STN demonstrates significant potential as a valuable resource for transit planners, researchers, and city planners. Its ability to evaluate transit system performance across various temporal and spatial scales provides crucial insights for improving public transportation networks.

By offering a comprehensive, flexible, and user-friendly platform for transit accessibility analysis, GTFS2STN contributes to the growing toolkit available to transportation professionals. As urban areas continue to grapple with issues of mobility and accessibility, tools like GTFS2STN will play an increasingly vital role in shaping efficient, equitable, and sustainable transit systems for the future.

\bibliographystyle{unsrtnat}
\bibliography{references}  






\end{document}